\documentclass[conference]{IEEEtran}
\IEEEoverridecommandlockouts
\usepackage{amsmath,amssymb,amsfonts}
\usepackage{algorithmic}
\usepackage{graphicx}
\usepackage{textcomp}
\usepackage{xcolor}

\usepackage{float}
\usepackage{hyperref} 

\renewcommand{\figurename}{Fig.} 

\def\BibTeX{{\rm B\kern-.05em{\sc i\kern-.025em b}\kern-.08em
    T\kern-.1667em\lower.7ex\hbox{E}\kern-.125emX}}
\begin{document}

\title{Energy-convergence trade off for the training of neural networks on bio-inspired hardware\\}

\author{
Nikhil Garg\textsuperscript{1,*}, 
Paul Uriarte Vicandi\textsuperscript{1,*}, 
Yanming Zhang\textsuperscript{1}, 
Alexandre Baigol\textsuperscript{1}, 
Donato Francesco Falcone\textsuperscript{2}, \\
Saketh Ram Mamidala\textsuperscript{2}, 
Bert Jan Offrein\textsuperscript{2}, 
Laura Bégon-Lours\textsuperscript{1}\\
\textsuperscript{1}Integrated Systems Laboratory, ETH Zurich, Switzerland, \\
\textsuperscript{2}IBM Research Europe, Zurich, Switzerland. 
Email:lbegon@ethz.ch
}

\maketitle

\vspace{-6mm}

{\footnotesize \textsuperscript{*}Equally contributed}

\begin{abstract}
The increasing deployment of wearable sensors and implantable devices is shifting AI processing demands to the extreme edge, necessitating ultra-low power for continuous operation. Inspired by the brain, emerging memristive devices promise to accelerate neural network training by eliminating costly data transfers between compute and memory. Though, balancing performance and energy efficiency remains a challenge. We investigate ferroelectric synaptic devices based on [HfO\textsubscript{2}/ZrO\textsubscript{2}] superlattices and feed their experimentally measured weight updates into hardware-aware neural network simulations. Across pulse widths from 20 ns to 0.2 ms, shorter pulses lower per-update energy but require more training epochs while still reducing total energy without sacrificing accuracy. Classification accuracy using plain stochastic gradient descent (SGD) is diminished compared to mixed-precision SGD. We analyze the causes and propose a ``symmetry point shifting'' technique, addressing asymmetric updates and restoring accuracy. These results highlight a trade-off among accuracy, convergence speed, and energy use, showing that short-pulse programming with tailored training significantly enhances on-chip learning efficiency.

\end{abstract}


\vspace{-6mm}

\section{Introduction}

Advances in wearable sensors and implantable devices \cite{wang2017flexible} are rapidly pushing artificial intelligence workloads to the extreme edge \cite{shi2016edge, rancea2024edge}, onto devices that must run from ultra-low-power envelopes, tolerate milli-joule-scale energy budgets per day, and minimize tissue heating \cite{vazquez2021compute}. Conventional CMOS processors struggle in this regime because transferring data between physically separated logic and memory dominates both energy consumption and latency \cite{horowitz20141}, even when aggressive voltage scaling is used. Inspired by the brain, where computation and memory are co-located, we therefore seek hardware paradigms such as neuromorphic engineering \cite{mead2020we, burr2017neuromorphic} and in-memory computing architectures \cite{sebastian2020memory} that diminish the boundary between storage and processing while remaining compatible with the tight thermal and power constraints of biomedical platforms.

In-memory computing with beyond-CMOS nanoscale devices \cite{theis2017end, mehonic2024roadmap} has emerged as a promising approach to accelerate compute-intensive operations in neural networks. Using fundamental physical laws (such as Ohm's and Kirchhoff's laws) to perform multiply-and-accumulate (MAC) operations directly in the memory array, these architectures can drastically reduce data movement and associated overheads. The synaptic weights, which connect one network layer to the next, can be implemented via memristive devices designed to mimic key synaptic functions: (i) storing analog or multi-level weights, and (ii) facilitating online learning. In addition to their nonvolatile retention and sub-femto-joule read energy \cite{covi2022challenges}, memristors offer multi-level programmability \cite{thomas2024versatile}, enabling on-chip adaptation \cite{wu2019preliminary} to continuously fluctuating physiological signals \cite{fang2022compact}.  Several material technologies (e.g., phase change, valence change) have demonstrated multi-state \cite{rao2023thousands} or analog behavior suitable for neural network training. Yet each programming event consumes pico- to nano-joules, and when scaled to billions of updates—this write energy can cause local heating and shorten battery life. Consequently, energy-efficient online programming strategies are pivotal for safe and sustainable operation. At the system level, software toolkits such as IBM’s \textit{AIHWKit} \cite{rasch2021flexible} and \textit{NeurosimMLP+} \cite{chen2017neurosim} can then simulate how these analog devices affect the end-to-end performance in real-world tasks.

In addition, these system-level simulation tools offer a powerful platform for investigating the significance of various figures of merit and to provide guidelines for the design and operation of synaptic weights. For bio-medical deployments, such studies must explicitly consider stringent power budgets and temperature constraints alongside accuracy.  At the device level, the voltage and duration of programming pulses may affect (i) the dynamic conductance range of the device, (ii) energy consumption during weight update, (iii) the symmetry of the potentiation and depression, and (iv) the number of conductance levels. At the neural network level, these device characteristics affect (i) the maximal accuracy and (ii) the number of epochs required to train the network (convergence speed). Consequently, finding effective design strategies that minimize the total amount of energy in online training without sacrificing accuracy remains a central challenge in the development of memristive in-memory computing systems. Previous studies have also addressed the issue of asymmetry in activation functions of neural networks \cite{liu2022weight} and memristive devices by specialized training algorithms, such as tiki-taka \cite{gokmen2020algorithm}, and proposed weight quantization schemes for memristive devices in \cite{kim2024study}.

In this study, we examine the total energy of the online training phase by measuring the trade-off between the convergence speed and the update energy per weight in in-house-fabricated ferroelectric synaptic devices \cite{begon2024back}. This metric directly relates to the thermal load and battery autonomy of an implantable AI coprocessor. Using a dedicated protocol to program multiple resistance states under varying pulse widths and amplitudes, we develop a device model in the AIHWKit simulation framework and evaluate stochastic gradient descent (SGD) algorithms within a multi-layer neural network. Our results show that, while shorter pulse widths require more training iterations, they drastically reduce overall training energy without compromising final accuracy on the MNIST dataset. Moreover, they allow tuning of the device’s inherently asymmetric switching behavior, further improving efficiency at the system level. We propose a weight initialization scheme that sets the reference of a differential pair \cite{prezioso2015training} to the weight symmetry point: "symmetry point shifting", and show that we can recover the precision of the neural network using plain SGD.

The paper is organized as follows. First, the fabrication of the synaptic weights is presented. Next, we outline the hardware-aware training strategies and detail the multi-pulse programming protocol used to fit the device model. We report the classification results on the MNIST digit classification benchmark, focusing on how the pulse width affects the convergence speed and energy usage. Finally, we address the switching polarity asymmetry, demonstrating that mixed-precision SGD or using shorter pulses with plain SGD can mitigate this behavior. We propose a symmetry-point shifting technique which, with plain SGD, can effectively restore balanced weight updates and preserve classification accuracy.

\section{Materials and Methods}

\subsection{Device fabrication}
\begin{figure}[!ht]
  \centering
  \includegraphics[width=0.45\textwidth]{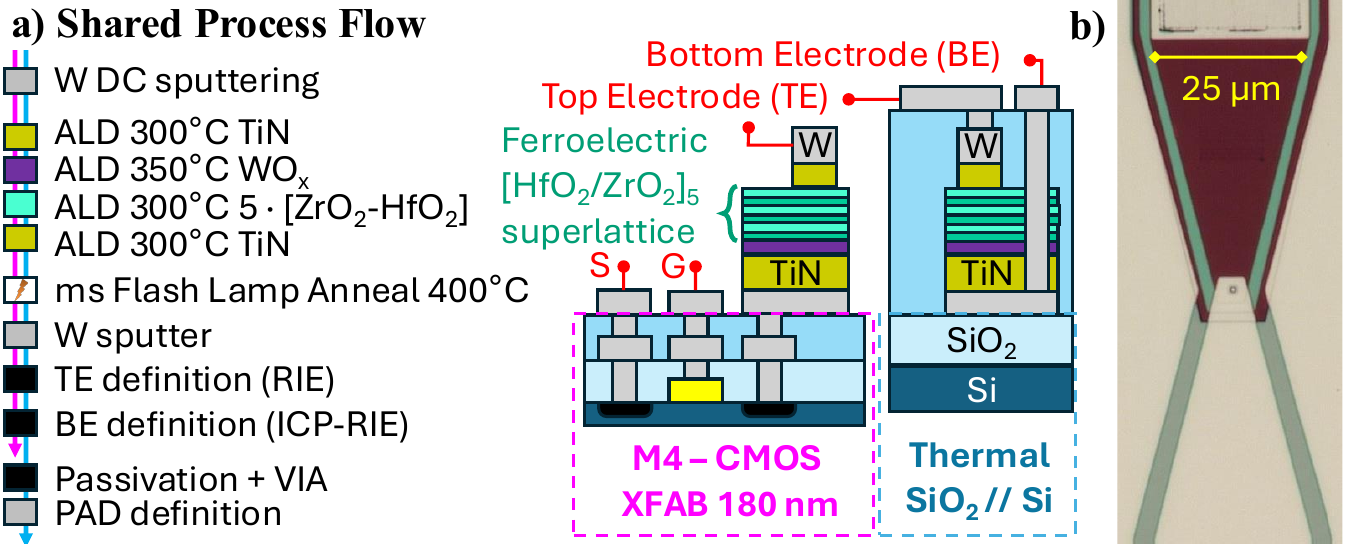}
  \caption{a) Shared process-flow for the fabrication of ferroelectric resistive devices integrated in the BEOL of XFAB 180 nm CMOS and on thermal SiO\textsubscript{2}//Si with schematic cross-sections. b) Microscope image of a 1 µm\textsuperscript{2} device. Scale bar is 25 µm.}
  \label{fig:fig1}
\end{figure}

\autoref{fig:fig1} illustrates the process flow and resulting device structures. The materials stack is identical to that of ref.\ \cite{begon2024back} which was integrated in the BEOL of CMOS technology (shared process-flow), but the substrate is a thermal SiO\textsubscript{2} (200 nm)//Si chip. Using Atomic Layer Deposition (ALD), 20 nm of TiN was deposited at 300\textdegree C, followed by 45 cycles of WO\textsubscript{x} at 360\textdegree C. Then five supercycles were deposited at 300\textdegree C alternating five cycles with tetrakis (ethylmethylamino) hafnium (IV) and O\textsubscript{2}, and ten cycles with bis (methylcyclopentadienyl) (methyl) (methoxy) zirconium (IV) and O\textsubscript{2}, forming a [HfO2/ZrO2] superlattice (SL).
Ten nanometers of TiN were deposited. Crystallization was performed with the millisecond flash lamp annealing technique: the sample was preheated at 375\textdegree C, followed by a 20 ms pulse of 90 J/cm\textsuperscript{2}.
A 50 nm thick W metal electrode was then sputtered. The top electrode was defined by e-beam lithography and reactive ion etching (RIE). The bottom electrode was defined by optical lithography and ion-beam etching of the SL, WO\textsubscript{x}, and TiN layers. 
A 100 nm thick SiO\textsubscript{2} passivation layer was then sputtered. Vias to the device’s top and bottom electrode contacts were defined by e-beam lithography. The SiO\textsubscript{2} layer was etched by RIE, and then the SL and the WO\textsubscript{x} were etched by ion beam etching. 100 nm of W was then sputtered. The metal lines were then defined by optical lithography and RIE.

\subsection{Neural network simulation}

\begin{figure}[!ht]
  \centering
  \includegraphics[width=0.495\textwidth]{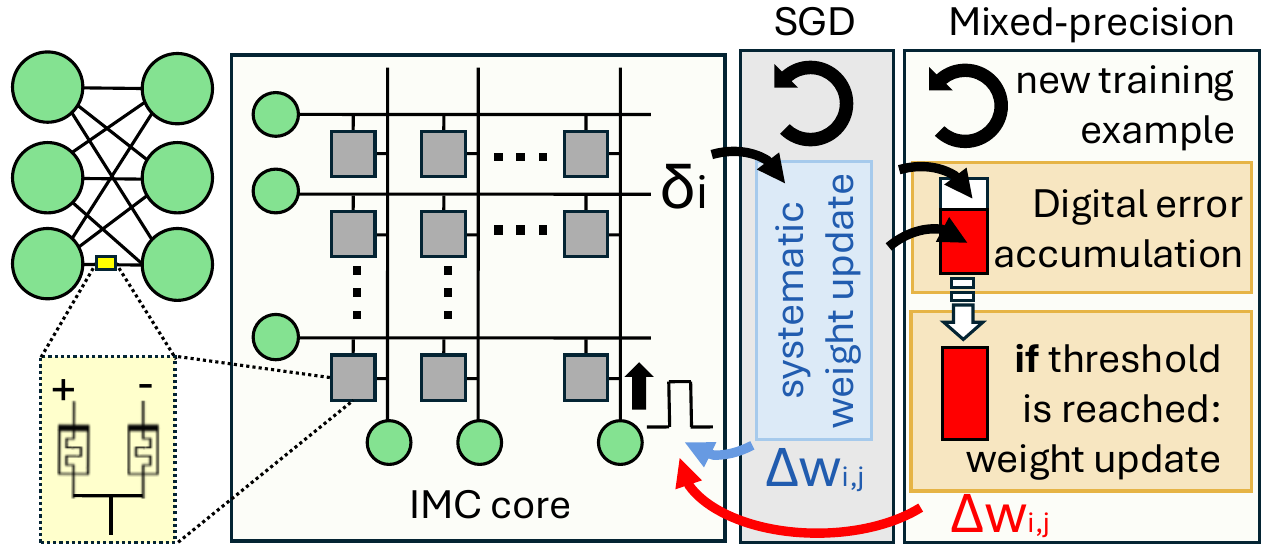}
  \caption{Training Neural Networks In-memory: synaptic weights are implemented using pairs of memristors. Training is either performed using plain SGD for systematic weight updates or using mixed-precision SGD, which digitally accumulates errors from multiple training samples and triggers weight updates once a predefined error threshold is reached.}
  \label{fig:fig2}
\end{figure}

The training methodology is illustrated in \autoref{fig:fig2}, where the synaptic weights are achieved using pairs of ferroelectric devices. Two primary training strategies are used: plain stochastic gradient descent (SGD) \cite{gokmen2016acceleration} and mixed precision SGD (MP-SGD). In MP-SGD \cite{nandakumar2020mixed}, digital errors accumulate until they exceed a specified threshold, upon which a weight update is triggered.

We evaluated our network using the MNIST \cite{lecun1998gradient} handwritten digit dataset, consisting of 28×28 grayscale images categorized into 10 classes (digits 0--9). The network topology is a four-layer deep neural network (784×256×28×10). The voltage setting and current reading were modeled through 8-bit analog-to-digital converters (ADCs) and digital-to-analog converters (DACs) from the AIHWkit. A piecewise fit was applied to model the characteristics of the ferroelectric memristive device. Specifically, we calculated the conductance gradient versus pulse number and used separate fifth-degree polynomial fits for the Long-Term Potentiation (LTP) and Long-Term Depression (LTD) gradients.

\section{Results}

\subsection{Electrical measurements}
\begin{figure}[!ht]
  \centering
  \includegraphics[width=0.49\textwidth]{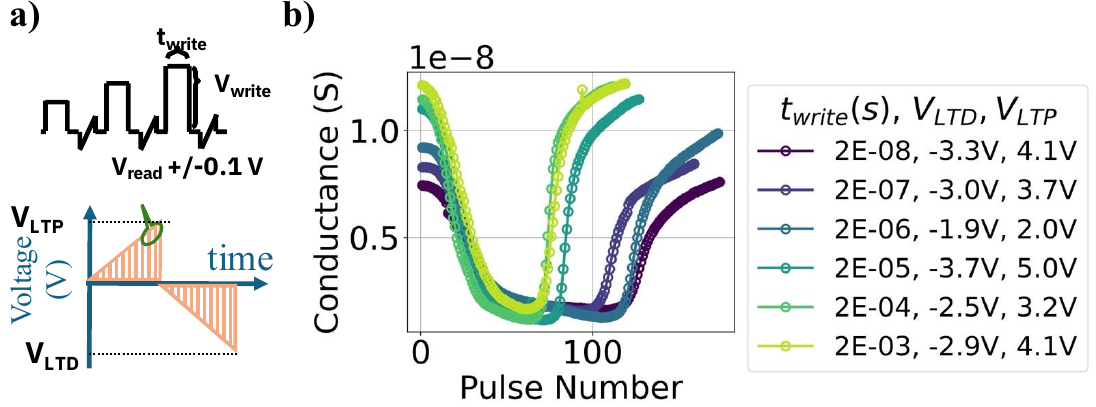}
  \caption{Long Term Potentiation and Depression of synaptic devices: a) A sequence of programming ``write'' pulses with increasing amplitude (V\textsubscript{write}) and fixed pulse width (t\textsubscript{write}) is applied. Each programming pulse is followed by a bipolar, non-destructive ``read'' pulse with fixed amplitude (V\textsubscript{read}=$\pm$0.1V). b) Evolution of synaptic weight (conductance) as a function of pulse number for different pulse widths: shorter t\textsubscript{write} require higher amplitudes.}
  \label{fig:fig3}
\end{figure}

The electrical characterization methodology is illustrated in \autoref{fig:fig3}. A Keysight B1500 semiconductor analyzer is used. The device undergoes a sequence of write pulses of fixed duration but increasing amplitude, and each write pulse is followed by a bipolar read pulse of $\pm$ 0.1V. As shown in \autoref{fig:fig3} (b), the conductance of the device (i.e., the synaptic weight) varies with the number of pulses applied for different pulse widths and amplitudes, demonstrating multilevel programmability.

\begin{figure}[!ht]
  \centering
  \includegraphics[width=0.49\textwidth]{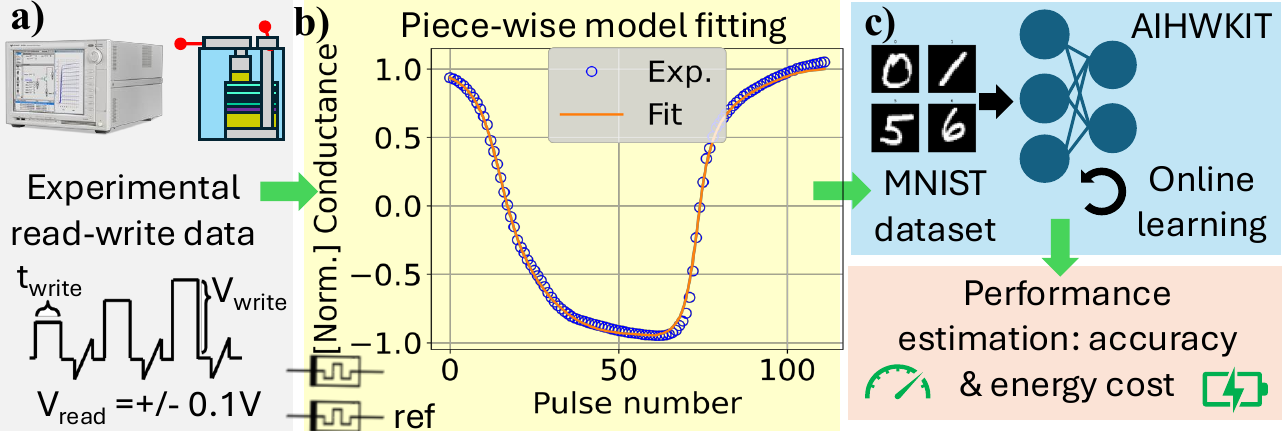}
  \caption{Hardware-aware neural network simulation framework: a) Devices are characterized using pulsed read-write measurement protocol. b) Experimental data is used to construct a piece-wise device model for neural network simulations. c) The neural network is trained on the MNIST handwritten digit dataset using the fitted device model, and performance metrics, including classification accuracy and estimated energy costs, are evaluated.}
  \label{fig:fig4}
\end{figure}

Experimental measurements (\autoref{fig:fig4}a) are used to build a piecewise model that captures the evolution of the conductance under various pulse conditions (\autoref{fig:fig4}b). Subsequently, this model is integrated into a hardware-aware neural network simulation, allowing for the evaluation of classification accuracy and energy consumption (\autoref{fig:fig4}c).

\subsection{Classification performance}
\begin{figure}[!ht]
  \centering
  \includegraphics[width=0.495\textwidth]{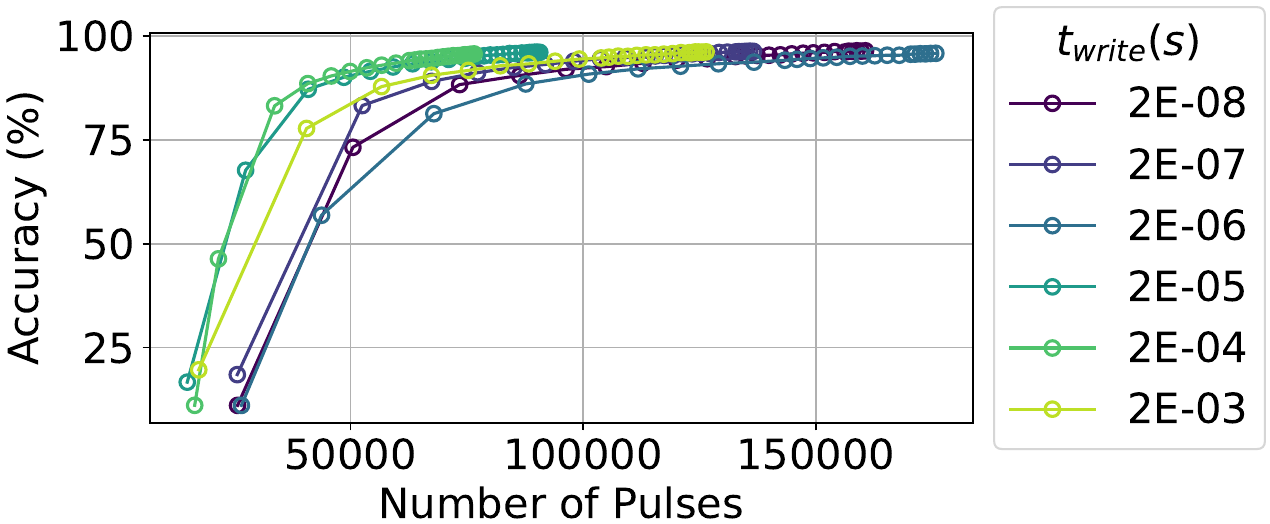}
  \caption{Network classification accuracy on the MNIST test dataset evaluated as a function of training epochs (data points), represented experimentally as the number of programming pulses. Performance evolution is compared across different programming pulse widths (t\textsubscript{write}).
  A weight update counter is implemented for the training with Mixed Precision SGD. After each epoch, the number of write pulses and the network accuracy is reported. For the same device, the shortest t\textsubscript{write} (\textgreater 1 µs) requires the most update pulses to converge to 95\%.
  }
  \label{fig:fig5}
\end{figure}

To evaluate classification performance, the trained network was tested on 10,000 previously unseen MNIST examples. The Mixed Precision SGD algorithm leads to the highest accuracy (95\%), comparable with that obtained by floating-point precision in purely software-based neural networks. A weight update counter is implemented to measure the number of updates after each epoch\footnote{https://github.com/NEO-ETHZ/CCMCC2025}. \autoref{fig:fig5} shows the classification accuracy over 25 training epochs, with pulse width (\textit{t\textsubscript{write}}) as a variable parameter.  Although shorter pulse widths initially show slower convergence, the accuracy converges to 95\% for all pulse widths.

\subsection{Trade-off: energy and convergence}

An upper bound on the weight-update energy cost is computed at each epoch using the pulse counter, the maximum amplitude for LTP (\textit{V\textsubscript{LTP}}), and \textit{t\textsubscript{write}} (see \autoref{fig:fig3}b): given a programming pulse with duration \textit{t\textsubscript{write}} and amplitude \textit{V\textsubscript{write}}, the energy dissipated per write operation is approximated by $E = I\cdot V_{LTP}\cdot t_{write}$, where \textit{I} is the current through the device during the pulse. To relate \textit{I} to \textit{V\textsubscript{write}}, we fit the experimental I-V characteristics with a modified Schottky emission model \cite{begon2024back}: the current density \( J \) is given by\cite{SimmonsMSE}:

$I = A \alpha T^{3/2} \frac{V}{d} \,\mu \exp \left( 
 -\frac{
     \Phi_B - \sqrt{\frac{V}{d} \,\frac{1}{4\pi \epsilon_0 \epsilon_r}}
 }{
     k_B T
 }
 \right)$

where \textit{A} = 4 µm\textsuperscript{2} is the device area, $\alpha$ = 3$\cdot$10\textsuperscript{-4} A$\cdot$s$\cdot$cm\textsuperscript{-3}$\cdot$K\textsuperscript{3/2} is a material-dependent prefactor, \textit{T} = 300 K is the absolute temperature, \textit{d} = 5 nm is the device thickness, \textit{µ} = 8.9$\cdot$10\textsuperscript{-3} cm\textsuperscript{2}$\cdot$s\textsuperscript{-1}$\cdot$V\textsuperscript{-1} is the carrier mobility, $\Phi_B$ = 0.19 eV is the Schottky barrier height, $\epsilon_0$ is the vacuum permittivity in consistent units, $\epsilon_r$ = 18.2 is the relative permittivity of the material, and \textit{k\textsubscript{B}} = 8.617$\cdot$10\textsuperscript{-5} eV$\cdot$K\textsuperscript{-1} is the Boltzmann constant.

\begin{figure}[!ht]
  \centering
  \includegraphics[width=0.45\textwidth]{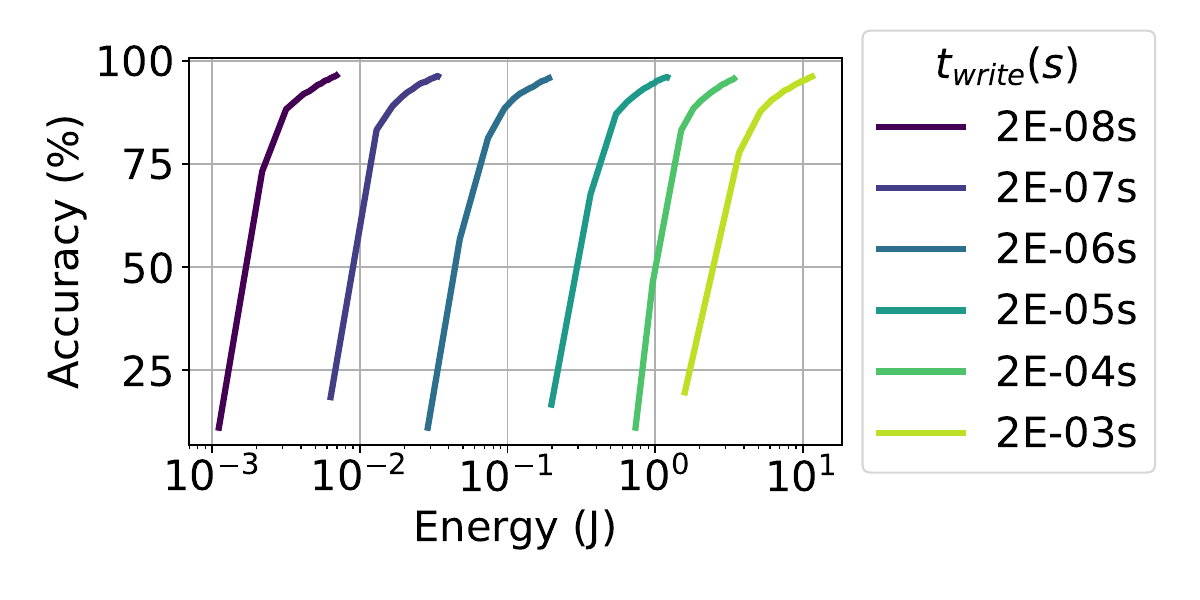}
  \caption{Estimated energy cost as a function of training epochs for different programming pulse widths (t\textsubscript{write}). Each data point represents a training epoch, showing the maximum estimated energy expenditure and corresponding network classification accuracy.
  }
  \label{fig:fig6}
\end{figure}

The energy-convergence trade-off is illustrated in \autoref{fig:fig6}, where the classification accuracy is plotted against the energy per training epoch for varying pulse widths (t\textsubscript{write}).  Although 20 ns pulses require more updates and a higher programming voltage, they nevertheless yield the lowest total energy cost to reach an accuracy of 95\%. 

\section{Discussion}

\begin{figure}[H]
  \centering
  \includegraphics[width=0.49\textwidth]{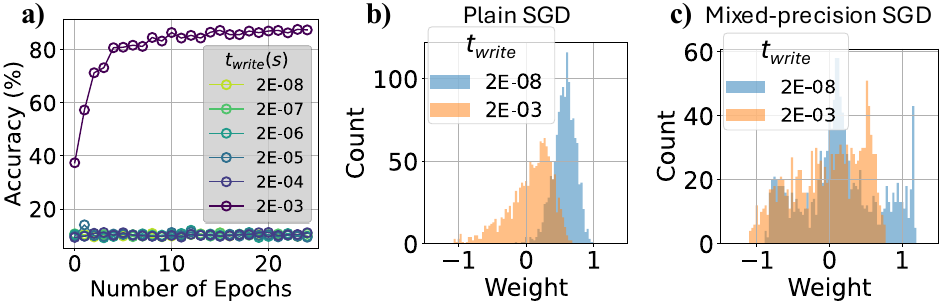}
  \caption{a) Network classification accuracy evaluated for plain SGD algorithm as a function of number of training epochs, for different programming pulse widths. Histograms of the final-layer network weights after training are compared at 2 pulse widths for: b) standard SGD, and c) mixed-precision SGD.
  }
  \label{fig:fig7}
\end{figure}

As we saw, mixed-precision SGD offers high neural network performance. However, it comes at the cost of increased hardware complexity, compared to the implementation of a plain SGD scheme, which allows for parallel weight update \cite{gokmen2016acceleration}. As shown in \autoref{fig:fig7}(a), plain SGD does not allow the training of the network for most device models studied previously, except for \textit{t\textsubscript{write}} = 2 ms. To investigate the poor classification performance of plain SGD, we examined the final-layer weight distributions under both plain SGD and mixed-precision SGD. \autoref{fig:fig7}(b) reveals that, with a plain SGD under 20 ns pulse widths, the weights cluster around approximately 0.59. This contrasts with the mixed-precision SGD case and plain SGD with 2 ms widths, both of which center around zero, as shown in \autoref{fig:fig7}(c).
\begin{figure}[H]
  \centering
  \includegraphics[width=0.45\textwidth]{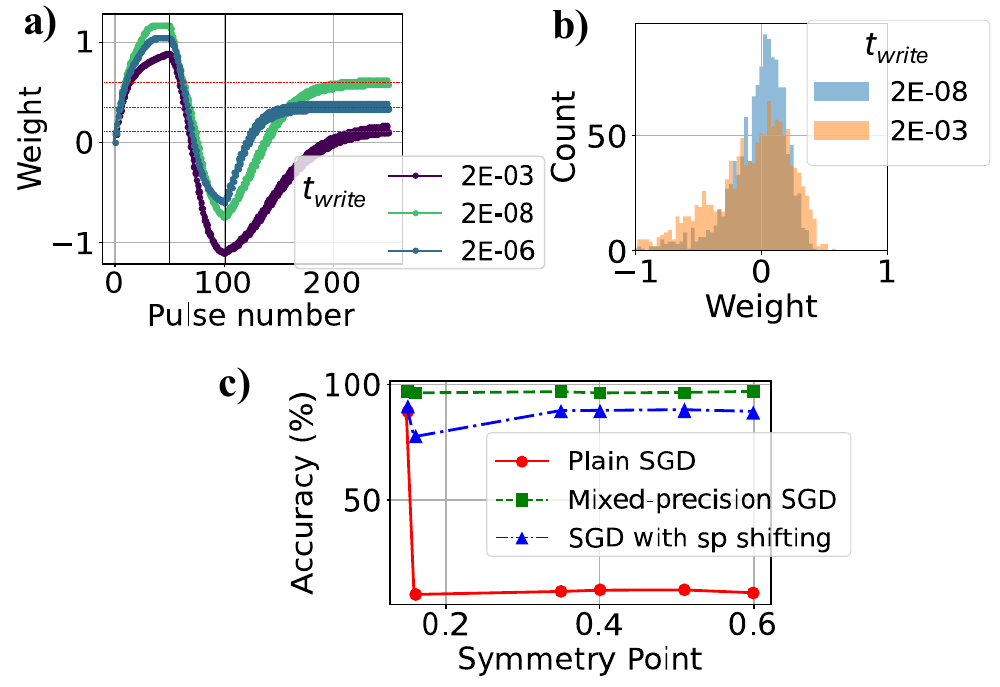}
  \caption{a) Conductance after 50 negative pulses, then 50 positive pulses, then alternating negative and positive pulses: depending on \textit{t\textsubscript{write}} the weight converges to a different symmetry point. b) Histograms of the final-layer network weights after training for SGD with symmetry-point shifting technique. c) Network classification accuracy evaluated for three training algorithms as a function of symmetry points corresponding to different programming pulse widths (\textit{t\textsubscript{write}}). }
  \label{fig:fig8}
\end{figure}

We explain this discrepancy through the concept of a ``symmetry point'', defined as the conductance level at which alternating positive and negative pulses yield no net change. The AIHWkit has a built-in module to determine it. By applying 50 pulses of one polarity, followed by 50 pulses of the opposite polarity, and then alternating pulses, we observe that different \textit{t\textsubscript{write}} values cause the device to stabilize at different symmetry points, as depicted in \autoref{fig:fig8}(a). 

To address this issue, we propose an approach in which one device in a differential synaptic pair is programmed to its symmetry point and held fixed while the other device undergoes standard SGD updates. A device-to-device variation of 0.05 is used \cite{begon2024back}. As illustrated in \autoref{fig:fig8} (b), this ``symmetry point shifting'' technique recenters the final weight distribution around zero. Consequently, it successfully recovers the classification accuracy lost under plain SGD alone. \autoref{fig:fig8}(c) compares accuracy across three training strategies: plain SGD, symmetry-point-shifted SGD, and mixed-precision SGD, highlighting that accounting for switching asymmetry restores most of the performance gap. 

\section{Conclusion}

In conclusion, we highlight several critical trade-offs for online learning with memristive devices. First, employing shorter pulse widths, which require higher programming voltages and often require more training epochs, significantly improves energy consumption estimates. Although this is promising and could potentially enhance the device's lifetime or endurance, it warrants further investigation. Second, mixed-precision SGD not only alleviates the challenges of asymmetric switching in memristive devices but also improves performance and reduces the number of weight updates, thereby extending the device lifetime and lowering energy consumption. However, the need for additional circuitry to handle gradient accumulation introduces hardware overhead. Third, we propose a ``symmetry point shifting'' technique, a subtle but crucial form of regularization that mitigates asymmetric behavior without requiring the same level of overhead as mixed-precision SGD. Although the present study employed the canonical MNIST image classification task, this serves as a first validation step toward real-time biosignal processing, such as electroencephalogram (EEG) \cite{chen2022global} and electrocardiogram (ECG) streams \cite{siontis2021artificial} in ultra-low-power wearable devices. Future work should therefore examine how the identified trade-offs scale to more complex network topologies and the temporal, multi-channel data characteristics of physiological signals, quantifying energy, latency, and thermal impact under realistic operating conditions.


\section*{Acknowledgment}

The authors acknowledge the Binnig and Rohrer Nanotechnology Center (BRNC). Research funded at ETH through the SNSF Starting Grant ROSUBIO (218438), by the Swiss State Secretariat for Education, Research, and Innovation (SERI) through
the SwissChips research project and through ViTFOX (101194368), IBM acknowledges co-funding of this work by the European Union under Grant agreement ID 101135398 (FIXIT) and 101135946 (CONCEPT).


\end{document}